\documentclass[aps,letter,nofootinbib,superscriptaddress]{revtex4} %eqsecnum,
\usepackage[dvips]{graphicx}

\usepackage{amssymb}
\usepackage{amsmath}
\usepackage{color}

\newcommand{\be}{\begin{equation}}
\newcommand{\ee}{\end{equation}}
\newcommand{\bea}{\begin{eqnarray}}
\newcommand{\eea}{\end{eqnarray}}
\newcommand{\eql}[1]{\label{eq:#1}}  
\newcommand{\ec}[1]{(\ref{eq:#1})}     
\newcommand{\refeq}[1]{equation~(\ref{eq:#1})}          
\newcommand{\refeqs}[2]{equations~(\ref{eq:#1})--(\ref{eq:#2})}          
          
\newcommand{\reffig}[1]{figure~\ref{fig:#1}}          
\newcommand{\refFig}[1]{Figure~\ref{fig:#1}}
\newcommand{\refapp}[1]{Appendix~\ref{app:#1}}
\newcommand{\vs}{\nonumber\\}       
\newcommand{\refsec}[1]{Section~\ref{sec:#1}}          

\renewcommand{\v}[1]{\vec{#1}}
\newcommand{\<}{\left\langle}
\renewcommand{\>}{\right\rangle}

\renewcommand{\k}{\kappa}
\renewcommand{\d}{\delta}
\newcommand{\D}{\Delta}

\newcommand{\ghat}{\hat{\gamma}}

\renewcommand{\l}{\ell}
\newcommand{\vth}{\v{\theta}}

\newcommand{\Wth}{W^\theta}

\newcommand{\dhat}{\hat{\d}}
\newcommand{\vl}{\v{\l}}

\newcommand{\nbar}{\bar{n}}
\newcommand{\nobs}{n_{\rm obs}}
\newcommand{\nfgobs}{n^{\rm fg}_{\rm obs}}
\newcommand{\dobs}{\d_{\rm obs}}

\newcommand{\bdobs}{\overline{\d_{\rm obs}}}
\newcommand{\bdobsfg}{\overline{\d^{\rm fg}_{\rm obs}}}
\newcommand{\bk}{\overline{\k}}
\newcommand{\bg}{\overline{\gamma}}
\newcommand{\tdobs}{\widetilde{\d_{\rm obs}}}
\newcommand{\udobs}{\overline{\d_{\rm obs}}}
\newcommand{\wdd}{\widehat{\dobs\dobs}}
\newcommand{\wddg}{\widehat{\d_g\d_g}}
\newcommand{\wddk}{\widehat{\k\k}}
\newcommand{\wddkg}{\widehat{\d_g\k}}

\newcommand{\wdh}[1]{\widehat{#1}}

\newcommand{\dobsfg}{\d^{\rm fg}_{\rm obs}}

\renewcommand{\a}{\alpha}

\renewcommand{\th}{\theta}
\newcommand{\g}{\gamma}
\newcommand{\vx}{\v{x}}
\newcommand{\NN}{\mathcal{N}}
\newcommand{\W}{\mathcal{W}}
\newcommand{\Wt}{\widetilde{\mathcal{W}}}

\begin{document}

\author{Fabian Schmidt}
\affiliation{Department of Astronomy \& Astrophysics, The University of
Chicago, Chicago, IL 60637}
\affiliation{Kavli Institute for Cosmological Physics, Chicago, IL 
60637}
\author{Eduardo Rozo}
\affiliation{CCAPP, The Ohio State University, Columbus, OH 43210}
\author{Scott Dodelson}
\affiliation{Center for Particle Astrophysics, Fermi National Accelerator 
Laboratory, Batavia, IL 60510}
\affiliation{Department of Astronomy \& Astrophysics, The University of
Chicago, Chicago, IL 60637}
\affiliation{Kavli Institute for Cosmological Physics, Chicago, IL 
60637}
\author{Lam Hui}
\affiliation{Institute for Strings, Cosmology, and Astroparticle Physics (ISCAP)}
\affiliation{Department of Physics, Columbia University, New York, NY 10027}
\author{Erin Sheldon}
\affiliation{Brookhaven National Laboratory, Upton, NY 11973}

\title{Lensing Bias in Cosmic Shear}

\begin{abstract}
Only galaxies bright enough and large enough to be unambiguously identified and measured are included in galaxy surveys used to estimate cosmic shear.  We 
demonstrate that because gravitational lensing can scatter galaxies across 
the brightness and size thresholds, cosmic shear experiments suffer from 
{\it lensing bias}. We calculate the effect on the shear power spectrum and 
show that -- unless corrected for -- it will lead analysts to cosmological 
parameters estimates that are biased at the 
$2-3\,\sigma$ level in DETF Stage III experiments, such as the Dark Energy 
Survey.  
\end{abstract}

%\date{\today}

\maketitle
%%%%%%%%%%%%%%%%%%%%%%%%%%%%%%%%%%%%%%%%%%%%%%%%%%%%%%%%%%%%%%%%%%%%%%%%%%%%%
%%%%%%%%%%%%%%%%%%%%%%%%%%%%%%%%%%%%%%%%%%%%%%%%%%%%%%%%%%%%%%%%%%%%%%%%%%%%%
\section{Introduction}

Weak gravitational lensing has emerged as a powerful tool to probe cosmological models. Current measurements~\cite{Hoekstra:2005cs,Massey:2007gh,Benjamin:2007ys} already constrain the amplitude of density perturbations in the universe and the total matter density. Future surveys are projected to have the power to constrain the most important parameters describing both dark energy~\cite{Albrecht:2006um,Albrecht:2009ct} and dark matter~\cite{Hu:1998az,Abazajian:2002ck,Ichiki:2008ye,Kitching:2008dp}.

While the largest uncertainties in these projections are experimental systematics, a lingering concern is our ability to make predictions for basic quantities such as the two-point function to sub-percent accuracy so that theoretical systematics will not be an issue. A number of higher order corrections to the two-point function have been considered: the Born correction, source-lens coupling, reduced shear, and lens-lens coupling~\cite{Bernardeau:1996un,Schneider:1997ge,Dodelson:2005ir}. Here we study another effect which contaminates the power spectrum at the same level as these~\cite{Hamana:2001kd}: {\it lensing bias}. 

Galaxies are selected in weak lensing surveys only if they are bright enough and large enough for their shapes to be adequately measured.  Lensing affects these criteria because galaxies too faint or small to make it into the catalog can be promoted into the sample if they are located in regions of large 
magnification. This effect is inevitable as it is only possible to cut
on {\it observed} sizes and magnitudes, and cannot be eliminated by imposing
brighter magnitude cuts.

  To appreciate the importance of this effect, consider a cartoon universe in which all galaxies are just a little too faint to be included in the survey.  In this case, only galaxies behind regions of large magnification would be included, so one would be able to estimate shear only behind foreground matter overdensities.  The ensuing shear map would be a map of clusters! Of course, reality is much more complicated than this toy example, and many galaxies will be in the survey by their own merits.  Moreover, the sky-dilution from lensing will compete with the effect we just described, so whether matter overdensities are over-sampled or under-sampled depends on the galaxy population.   Nevertheless, it is clear 
that the sampling of the cosmic shear field from a typical galaxy survey will 
almost always be biased. In this paper, we derive this lensing bias and study
its effect on the shear power spectrum. We also discuss how it affects
other shear observables.

In \refsec{lbias}, we present and discuss the leading lensing bias corrections.
\refsec{results} then calculates the correction to the shear power spectrum,
while other shear observables are discussed in \refsec{other}. We conclude
in \refsec{concl}. The appendices contain a rigorous derivation of
the leading and higher order correction terms, and discuss why the higher
order terms can be neglected for the purposes of near-future surveys.

%%%%%%%%%%%%%%%%%%%%%%%%%%%%%%%%%%%%%%%%%%%%%%%%%%%%%%%%%%%%%%%%%%%%%%%%%%%%%
%%%%%%%%%%%%%%%%%%%%%%%%%%%%%%%%%%%%%%%%%%%%%%%%%%%%%%%%%%%%%%%%%%%%%%%%%%%%%
\section{Lensing bias and Cosmic Shear}
\label{sec:lbias}

This section describes the leading order lensing bias effects on cosmic
shear. For a rigorous derivation and treatment of higher order terms,
see Appendices \ref{app:exact} and \ref{app:projection}.
Let us consider a survey of solid angle $\Delta\Omega$ with
$N_{\rm tot}$ observed galaxies in total, so that the observed average number
density is $\nbar = N_{\rm tot}/\Delta\Omega$. 

To first order, the {\it observed} galaxy overdensity $\dobs$ is given in 
terms of the intrinsic galaxy overdensity $\d_g$ and the convergence $\k$ 
by \cite{paperI}:
\be
\dobs(\vth) = \d_g(\vth) + q\:\k(\vth), %\quad q = 2\beta_f + \beta_r - 2,
\label{eq:dobs}
\ee
where $q=2\beta_f+\beta_r-2$,
and $\beta_f$ and $\beta_r$ are the logarithmic slopes of the flux and size 
distributions,
\be
\beta_f = -\frac{\partial \ln n_{\rm obs}}{\partial \ln f}\Bigg\vert_{\stackrel{f=f_{\rm min}}{r=r_{\rm min}}}; \ \  \beta_r = -\frac{\partial \ln n_{\rm obs}}{\partial \ln r}\Bigg\vert_{\stackrel{f=f_{\rm min}}{r=r_{\rm min}}}.
\label{eq:beta}
\ee
In the following, $\dobs$ will always stand for the background galaxies
whose shear is measured, while foreground galaxy overdensities we correlate
with will be denoted with $\dobsfg$ for clarity. 

In the weak lensing limit, cosmic shear can be described by a spin-2
field with two independent components, defined relative to fixed coordinate
axes $x,y$ ($\rightarrow \g_1$, $\g_2$), or with respect to the separation 
vector $\vth$ ($\rightarrow \g_t$, $\g_\times$). In the following, we let $\g_a$ and $\g_b$ stand for
either of these decompositions. We work in the flat sky approximation throughout,
denoting positions on the sky with $\v{x}$. 

Let $\g_a(i)$ be the shear component $a$ measured from galaxy $i$.
The standard estimator for shear correlation functions $\xi_{ab} \equiv \<\g_a\g_b\>$ is given by (e.g., \cite{MunshiEtal}):
\be
\hat\xi_{ab}(\th) = \frac{1}{N} \sum_{ij} \Wth(i,j)\, w(i)w(j) \gamma_a(i)\gamma_b(j), %\quad N = \sum_{ij} \Wth(i,j)\,w(i) w(j),
\label{eq:xiest}
\ee
where the sum runs over all pairs of galaxies $i, j$, and the normalization is
given by:
\be
N = \sum_{ij} \Wth(i,j)\,w(i) w(j),
\ee
where $w(i)$ is the weight assigned to galaxy $i$, and the window function 
$\Wth$ picks out galaxies separated 
by $\theta-d\th \leq |\vx_i-\vx_j| <\theta+ d\th$.   For the remainder
of the paper, we will set all weights $w(i)=1$, assuming that they
are determined by measurement errors and intrinsic ellipticities \cite{BJ02},
and are therefore uncorrelated with the cosmological signal. We also
assume that the shape noise is uncorrelated with the density field.

In \refapp{pixelest}, we consider pixel-based estimators. As shown there, 
pixel-based estimators for the shear
correlation function are subject to a very similar bias to the galaxy pair-based
estimator above, provided that each pixel is weighted by inverse variance.

We wish to take the expectation value of \refeq{xiest}.  To do so, we 
partition the survey volume into infinitesimal cells of equal solid angle $d\Omega$
so that the number of galaxies $\nobs(i)d\Omega$ in cell $i$ is either 0 or 1 
for all cells.  Given this partition, we can express \refeq{xiest} as:
\be
\hat\xi_{ab}(\th) =  \frac{1}{N} \sum_{ij}\Wth(i,j)\,\nobs(i) d\Omega\, \gamma_a(i)\,\nobs(j) d\Omega\, \gamma_b(j) 
\label{eq:xihat1}
\ee
where the sum is now over all cells.  The normalization $N$ can be similarly 
rewritten.  Now, we have that $\nobs(i)=\nbar\,(1+\dobs(i))$ where 
$\dobs(i)$ is the fluctuation in the galaxy density field in cell $i$.     
Inserting these expressions 
into the above equation and taking the expectation value in the continuum 
limit, we find (see \refapp{exact} for details): 
\be
\<\hat\xi_{ab}(\th)\> =  \< \frac{1}{\NN} [1+\dobs(1)]\g_a(1)\:
[1+\dobs(2)]\g_b(2) \>,
\label{eq:xihat2}
\ee
where we have denoted two positions separated by $\th$ on the sky with `1' 
and `2'. The quantity $\NN$ is defined in \refapp{exact} and comes from 
the normalization by the observed
number of galaxy pairs. 

The important point to note here is the fact that the non-uniform sampling 
of source galaxies through $1+\dobs$ makes the estimator in \refeq{xiest} 
sensitive not only to the shear but also to the source galaxy 
overdensity {\it and} the lensing magnification.  Operationally
(with the caveat of higher-order corrections), the estimator 
in \refeq{xiest} replaces the true shear $\gamma(i)$ by an 
``observed'' shear:
\begin{equation}
\g_a^{\rm obs} \rightarrow \g_a (1+\dobs) = \g_a\left( 1+ \delta_g  + q\kappa\right).
\label{eq:gscaling}
\end{equation}
Hence, by expansion of \refeq{xihat2} we obtain the leading corrections:
\bea
\<\hat\xi_{ab}(\th)\> &=& \<\g_a(1)\g_b(2)\> \nonumber\\
&+& \<\,[\d_g(1)+q\,\k(1)]\g_a(1)\g_b(2)\> 
  + \<\g_a(1)\,[\d_g(2) + q\,\k(2)]\g_b(2)\> \label{eq:xiexp}
\eea
The correction terms are of two kinds: one involves correlations of
$\d_g\g_a$, i.e. intrinsic overdensities of background galaxies with shear by 
mass fluctuations in the foreground.
For a sufficiently narrow redshift distribution of source galaxies, 
this source-lens clustering is negligible, since the distribution of
sources and lenses do not overlap in this case. In case of the lensing
skewness and kurtosis, it was shown that the effect is small if the
width of the source redshift distribution is less than 0.15 \cite{Bernardeau98}.
Hence, if photometric redshifts are available, source-lens
clustering can be avoided.
We will not further consider source-lens clustering
in the main part of the paper.

The second type of correction in \refeq{xiexp} is due to magnification
and size bias and is of the form $q\cdot\k\g_a$. These corrections
can be significant, since they correlate the shear field with the {\it same} 
foreground lensing field. 
It is worth noting that the leading lensing bias corrections are of exactly the same form as the reduced shear correction \cite{Dodelson:2005ir,DodelsonZhang,Shapiro}: there, $\g \rightarrow \g (1+\k)$  perturbatively, whereas here, we have $\g \rightarrow \g (1+q\,\k)$. Hence, reduced shear corrections and lensing bias corrections should be considered jointly. The main difference is that the size of lensing bias corrections depends on the background galaxy sample via the parameter $q$. From now on, we consider both effects simultaneously, so that:
\begin{equation}
\g_a^{\rm obs} \rightarrow  \g_a\left [ 1+ (1+ q)\kappa \right ].
\label{eq:gammaobs}
\end{equation}

Note that the normalization $\NN$ in \refeq{xihat2}
is relevant in canceling some higher-order
terms. In particular, one might wonder whether the term:
\be
\< \d_g(1)\d_g(2)\> \<\g_a(1)\g_b(2)\>,
\ee
which appears in the expansion of \refeq{xihat2} might be a significant 
contribution. This term is however canceled through $\NN$, since the
shear estimator is normalized to the number of observed galaxy pairs used in
the measurement (see also \refapp{projection}).
The contributing higher-order terms due to lensing bias which we neglected 
in \refeq{xiexp} involve the shear 4-point 
functions. We discuss these terms in \refapp{projection} and find that
they are suppressed by roughly two orders of magnitude
with respect to the cubic terms, i.e. they entail corrections at the level of 
$\mathcal{O}(10^{-4})$ of the shear power spectrum.
Corrections of this magnitude are not expected to be of interest in the
foreseeable future, hence we neglect them for the remainder of the paper.

%%%%%%%%%%%%%%%%%%%%%%%%%%%%%%%%%%%%%%%%%%%%%%%%%%%%%%%%%%%%%%%%%%%%%%%%%%%%%
\section{Impact on the Power Spectrum}
\label{sec:results}

In this section, we present the results of a calculation of the leading
magnification effects on the shear auto-correlation, \refeq{xiexp}.
Specifically, we will consider 
$\xi_\g \equiv \<\g\, \g^*\> = \<\g_1\g_1\> + \<\g_2\g_2\>$
for background galaxies at a fixed redshift of $z_s=1$.
The cubic corrections involve three-point functions of shear and convergence
It is much more convenient to calculate these in Fourier space where, in the
absence of B modes, the complex shear is related to the convergence as:
\be
\g(\vl) = e^{2i\phi_\l}\:\k(\vl).
\label{eq:gk}
\ee
Here, $\phi_\l$ is the angle of the $\vl$ vector with the $\vx$-axis of
the coordinate system. Then, the shear power coefficients 
$C^\k(\l)$ are defined as:
\be
\<\g(\vl)\g^*(\vl')\> = \<\k(\vl)\k(-\vl')\> = (2\pi)^2 \d_D(\vl-\vl') C^\k(\l),
\label{eq:g2pt}
\ee
and their relation to the real-space correlation function is given by:
\be
C^\k(\l) = \int d^2\th\:\xi_\g(\th) e^{-i\,\vl\cdot\v{\th}}.
\ee
The calculation then proceeds exactly as
in the case of the reduced shear correction \cite{Dodelson:2005ir,DodelsonZhang,Shapiro}. In Fourier space, the multiplication in \refeq{gammaobs} turns into
a convolution:
\be
(\k\g)(\vl) = \int\frac{d^2\l_1}{(2\pi)^2}
\g(\vl_1)\k(\vl-\vl_1) = \int\frac{d^2\l_1}{(2\pi)^2}
e^{i 2\phi_{l_1}} \k(\vl_1)\k(\vl-\vl_1),
\ee
where we have used \refeq{gk}. Then, the leading correction to the two-point
correlator \refeq{g2pt} is given by:
\be
\d \<\g(\vl)\g^*(\vl')\> = 2(1+q) \int\frac{d^2\l_1}{(2\pi)^2}
e^{i 2\phi_{l_1}} e^{-i 2\phi_{l'}} 
\< \k(\vl_1)\k(\vl-\vl_1)\k(-\vl')\>,
\ee
\begin{figure}[ht!]
\begin{center}
\includegraphics[width=.48\textwidth]{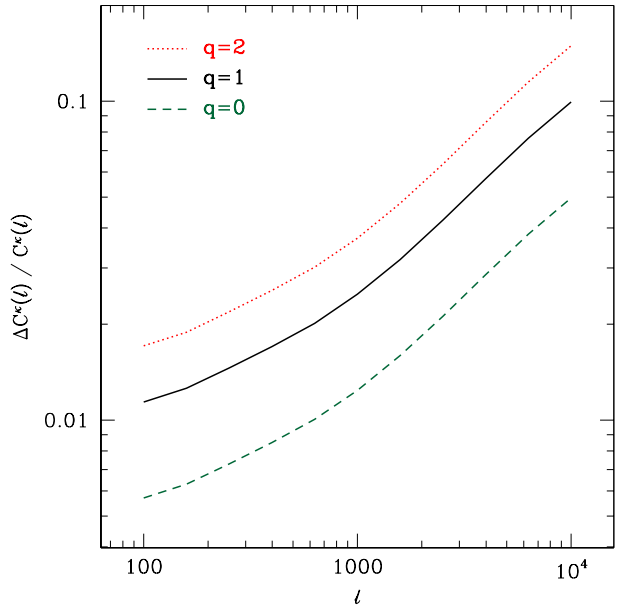}
\end{center}
\caption{Relative size of the combined lensing bias and reduced shear correction
$\Delta C^\k(\l)$ on the shear power, for different values of the flux/size
count slope $q$. The curve for $q=0$ only shows the reduced shear correction.
A single source redshift $z_s=1$ was assumed.
\label{fig:dCkappa}}
\end{figure}

where the factor of 2 comes from the two permutations.
Using the definition of the convergence bispectrum,
\be
\<\k(\vl_1)\k(\vl_2)\k(\vl_3)\> = (2\pi)^2 \d(\vl_1+\vl_2+\vl_3)
B^\k(\l_1,\l_2,\l_3),
\ee
we obtain the correction to the shear power spectrum:
\be
\D C^\k(\l) \equiv C^\k_{\rm corr}(\l)-C^\k(\l) = 2 (1+q)\int\frac{d^2\l_1}{(2\pi)^2}
e^{2i(\phi_{\l_1}-\phi_\l)} B^\k(\vl_1,\vl-\vl_1,-\vl).
\label{eq:dCkappa1}
\ee
The imaginary part of \refeq{dCkappa1} vanishes, signaling that no B modes
are produced by these 3-point terms (see Appendix~\ref{app:b} for a treatment of the B-modes induced by the 4-point terms). The remaining real part is:
\be
\D C^\k(\l) = 2 (1+q)\int\frac{d^2\l_1}{(2\pi)^2}
\cos 2\phi_{\l_1} B^\k(\vl_1,\vl-\vl_1,-\vl)
\label{eq:dCkappa}
\ee
Here, we have set $\phi_\l=0$ without loss of generality. The prefactor
$1+q$ in \refeq{dCkappa} sums up the reduced shear and lensing bias 
corrections.

\begin{figure}[ht!]
\begin{center}
\includegraphics[width=.48\textwidth]{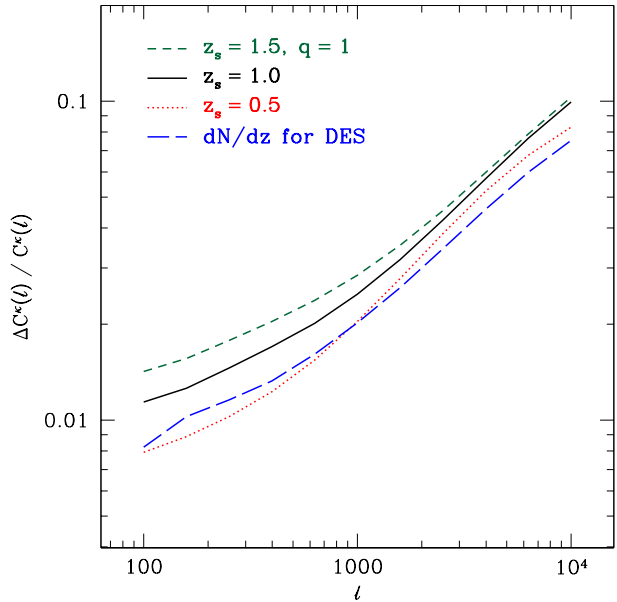}
\end{center}
\caption{Same as \reffig{dCkappa}, except for different source redshift
distributions: the lines for $z=0.5$, $z=1.0$ and $z=1.5$ show the result
for single source redshifts, while the blue dashed line shows the result
for a broad redshift distribution $z=0-1.4$ expected for the full DES 
galaxy sample \cite{Annis}.
\label{fig:dCkappa-z}}
\end{figure}

To estimate $\Delta C^\k(l)$, we adopt a flat $\Lambda$CDM cosmology with 
parameters given by
$h=0.7$, $\Omega_m=1-\Omega_\Lambda=0.28$, $n_s=0.96$, $\sigma_8 = 0.85$.
We use \refeqs{Ckappa}{Bkappa} in \refapp{projection} 
relating the shear power and bispectrum
to the matter power spectrum and bispectrum.
Further, we use the non-linear matter power spectrum according to \cite{SmithEtal} 
together with the bispectrum fitting formula from \cite{SC}. 
Our calculation of the reduced shear correction
agrees with that of \cite{Dodelson:2005ir,Shapiro} where it was shown to
match the results of ray-tracing through N-body simulations.

Figure \ref{fig:dCkappa} shows the relative magnitude of the cubic correction
$\Delta C^\k(\l)/C^\k(\l)$ from \refeq{dCkappa} for 
a range of $q$ values from 0 to 2, for a fixed source redshift of $z_s=1$.  
In \citet{paperI}, we consider a galaxy sample similar to the one 
expected for the Dark Energy Survey
\citep[DES,][]{DES05}. Measuring the slope of the galaxy size and
magnitude distributions according to \refeq{beta} for 
a range of magnitude and size cut values, we obtain $q\approx 1-2$
(see \cite{paperI} for details).

At $\l\sim 1000$, the cubic correction term reaches about 4\% for $q=2$,
which is larger than what one might naively expect
from perturbation theory.  This is because of a much larger weighting
of low-redshift contributions in the weak lensing bispectrum when
compared to the power spectrum (\refapp{projection}): the correction 
\refeq{dCkappa} is
enhanced by the more strongly non-linear matter distribution at low $z$.

We show the effect of varying source redshifts in \reffig{dCkappa-z}.
The relative correction to $C^\k(\l)$ increases with redshift, although
the $z$-dependence is quite weak. We also consider a very broad
galaxy redshift distribution $dN/dz$ expected for the full DES 
galaxy sample \cite{Annis}, spanning redshifts from 0 to 1.4.
Even in this case, the magnitude
of the effect is not affected significantly.

%%%%%%%%%%%%%%%%%%%%%%%%%%%%%%%%%%%%%%%%%%%%%%%%%%%%%%%%%%%%%%%%%%%%%%%%%%%%%
%%%%%%%%%%%%%%%%%%%%%%%%%%%%%%%%%%%%%%%%%%%%%%%%%%%%%%%%%%%%%%%%%%%%%%%%%%%%%
\section{Lensing bias corrections to other shear observables}
\label{sec:other}

While we chose the cosmic shear power spectrum as a representative example
to illustrate the magnification and size bias effects, it is worth considering
briefly the effects on other observables. In the following, we assume
values of $q=1-2$ as typical.

\subsection{Mean Shear}  

Lensing bias (and reduced shear) do not affect the mean of the shear, or
convergence if the convergence is estimated from the shear. That is, 
one might worry that, since regions of large $\kappa$ are preferentially 
selected, the average value of the shear in all pixels in a survey might be 
pushed to a non-zero value. This is not the case. Consider the estimator
for the convergence:
\begin{equation}
\kappa^{\rm obs}(\v{x})=\int \frac{d^2\l}{(2\pi)^2} e^{i\vl\cdot\v{x}} \int d^2 x' e^{-i\vl\cdot\v{x}'} \left[ \cos(2\phi_\l) \gamma^{\rm obs}_1(\v{x}')
+\sin(2\phi_\l) \gamma^{\rm obs}_2(\v{x}')\right]\eql{mean}
\end{equation}
where $\phi_\l$ is the angle between $\vl$ and the $x$-axis, and $\gamma^{\rm obs}_a(\v{x})$ are the measured ellipticities. We have argued that inevitably 
\begin{equation}
\gamma^{\rm obs}_a(\v{x}) \rightarrow \gamma_a(\v{x})\left[1+\delta_{\rm obs}\right]
\end{equation}
so contains quadratic terms such as $\gamma_a(\v{x})\kappa(\v{x})$. Symmetry though dictates that the means of these terms, $\langle \gamma_a(\v{x}')\kappa(\v{x}') \rangle$, in \refeq{mean} vanish: $\gamma_1$ is just as likely to be positive as negative so $\gamma_1\kappa$ averages to zero. The mean therefore of any
{\it linear} combination of the shear components remains zero in the presence of the corrections considered here.

\subsection{B-Modes}
\label{sec:Bmodes}

Cooray and Hu~\cite{Cooray:2002mj} pointed out that corrections to the Born approximation inevitably lead to non-zero B-modes. Lensing bias (and reduced shear) also produce B-modes; the Gaussian contribution to the spectrum is (see Appendix \ref{app:b}):
\begin{equation}
C^B(\l) = (1+q)^2 \int\frac{d^2\l'}{(2\pi)^2}
\sin(2\phi_{\l'}) 
C^\kappa(\l') C^\kappa(\vert\vl-\vl'\vert) \left[ \sin(2\phi_{\l'})+ \sin(2\phi_{\vl'-\vl})\right].
\end{equation}
Apart from geometric factors and the prefactor, this is of order 
$l^2C^\kappa(l) \sim 10^{-4}$  smaller than the E-mode spectrum, 
in qualitative agreement with the terms analyzed in \cite{Cooray:2002mj}. 
While the factor of $(1+q)^2$ could provide a boost of order 10 to this 
B-mode power spectrum, the amplitude is still likely too small to be detected 
in upcoming surveys. Therefore, B-modes will continue to serve as excellent 
checks of systematic effects. Note that on very small scales, the 
non-Gaussian contribution from the trispectrum
might be significant. A calculation of this contribution is much more involved
and is left for future work.

%%%%%%%%%%%%%%%%%%%%%%%%%%%%%%%%%%%%%%%%%%%%%%%%%%%%%%%%%%%%%%%%%%%%%%%%%%
\subsection{Galaxy-shear correlation}
\label{sec:galshear}

Following an argument analogous to the one presented for the shear-shear correlation function, we can derive the impact of lensing bias on the galaxy-shear correlation function.  The equivalent of \refeq{xihat2} for the correlation of a background shear $\g_a$ with foreground galaxies is: 
\be
\<\hat\xi_{ga}(\th)\> = \left \langle \frac{1}{\NN'}\dobsfg(1)\:
[1+\dobs(2)]\g_a(2) \right \rangle,
\label{eq:xihatgk}
\ee
where $\NN'$ again is from the normalizing denominator in the estimator 
defined in \refapp{exact}.
Expanding \refeq{xihatgk}, we obtain the 
following corrections to the galaxy-shear correlation:
\bea
\<\hat\xi_{ga}(\th)\> &=& \<\d^{\rm fg}_g(1)\g_a(2)\> + q^{\rm fg} \<\k^{\rm fg}(1)\g_a(2)\> \nonumber\\
&+& \<\,[\d^{\rm fg}_g(1) + q^{\rm fg}\,\k^{\rm fg}(1)]\,\d_g(2)\g_a(2)\> \nonumber\\
&+& q \<\,[\d^{\rm fg}_g(1) + q^{\rm fg}\,\k^{\rm fg}(1)]\,\k(2)\g_a(2)\>
\label{eq:xiexpgk}
\eea
Here, $q^{\rm fg}$ and $\k^{\rm fg}$ denote $q$ and convergence for the 
foreground galaxies. The first line of \refeq{xiexpgk} shows the lowest order 
contributions,
including the lensing bias of foreground galaxies \cite{ZiourHui,Bernstein08},
while the second line shows the corrections due to source-lens clustering
which we again assume to be small. Finally, the third line shows the 
contributions due to lensing effects on the background shear. These
are again similar to the corresponding reduced shear correction,
apart from the factor of $q$.
In total, we expect
the lensing bias effects on the galaxy-shear correlation to be
of similar size as those on the shear autocorrelation (\refsec{results}),
and to scale similarly with redshift.

\subsection{Shear tomography} 
\label{sec:tomo}

Measuring shear correlations between different
redshift slices allows for precise constraints on the expansion history
of the Universe and the dark energy equation of state, $w$. Using the Fisher matrix
technique, Shapiro \cite{Shapiro}
estimated the dark energy parameter biases incurred when neglecting 
the reduced shear correction. Since the change to the shear-shear power spectrum
due to reduced shear and lensing bias scales as $1+q$, we expect the corresponding
parameter biases to increase by a factor of $2-3$ for $q=1-2$ if these effects
are neglected.
For example, for a DES-like survey (DETF Stage-III), we expect a biasing
of $w$ at the $2-3\,\sigma$ level for a flat $w$CDM model.

\subsection{Shear variance and aperture mass}
\label{sec:variance}

Apart from the correlation function,
shear auto-correlations are often measured
in terms of top-hat variance $\gamma_\theta$ and aperture mass 
$M_{\rm ap}(\theta)$ (e.g., \cite{MunshiEtal,FuEtal}).
These estimators use window functions which weight angular scales in
different ways. White \cite{White05} showed that the reduced shear
correction has a $\sim 12$\% effect on the top-hat shear variance
even on large scales. This is because small scales contribute strongly
to the shear variance. Following the results of the previous section, 
we expect this effect to be amplified by $(1+q)$ to a total of
$\sim 24-36$\% when including both reduced shear and lensing bias. 
In case of the aperture mass, angular scales much smaller than the
filter scale are downweighted, so that
the effect on the aperture mass variance is smaller;
we expect a $\sim 10-25$\% correction for $\theta \lesssim 4\,$arcmin.

%%%%%%%%%%%%%%%%%%%%%%%%%%%%%%%%%%%%%%%%%%%%%%%%%%%%%%%%%%%%%%%%%%%%%%%%%%%%%
%%%%%%%%%%%%%%%%%%%%%%%%%%%%%%%%%%%%%%%%%%%%%%%%%%%%%%%%%%%%%%%%%%%%%%%%%%%%%
\section{Conclusions}
\label{sec:concl}

Estimates of cosmic shear suffer from lensing bias: the way one selects 
galaxies to estimate shear is correlated with the shear field itself.  
This correlation reflects the fact that cosmic shear and 
magnification are due to the same foreground matter along the line of sight,
and magnification and size bias can scatter source galaxies into or out of 
the galaxy catalog.

Lensing bias needs to be understood if lensing is to be used as a precision probe of the
dark sector. We estimated that neglecting lensing bias and its cousin reduced shear when interpreting the results of a DETF Stage III cosmic shear experiment
such as DES will
lead to estimates of the dark energy equation of state which differ from the true value by 2-3
statistical standard deviations. Thus, this is an important 
systematic that needs
to be addressed. In fact, lensing bias is likely to pollute other lensing 
measures even
more severely: piggy-backing on the calculation of \cite{White05} for reduced shear alone, we estimate 
that lensing bias + reduced shear will affect shear variance and 
aperture mass variance at the $20-30$\% level.
This could well be of importance to weak-lensing selected clusters, since
cluster finding and mass estimates from weak lensing are based on 
estimators similar to aperture mass or top hat variance \cite{HamanaEtal2004,MaturiEtal2005}.
Lensing bias is also likely to be the most significant source of cosmological 
B-modes in the shear field, at roughly $\sim 10^{-3}$ 
of the E-modes on small scales for the Gaussian contribution.

Correcting for these biases should not be too difficult: the perturbative 
calculation presented here
has been shown to agree with simulations~\cite{Dodelson:2005ir}. While better 
calibration is needed, this is clearly a
solvable problem, especially on large scales where baryons are not a factor. 
Moreover, one can imagine calibrating from the data itself by varying size 
and magnitude cuts to isolate $q$. Indeed, one possible application 
of lensing bias is as a calibrator for multiplicative and additive shear 
errors. We plan to explore this possibility in future work.

\acknowledgments
We would like to thank Chaz Shapiro for helping us in cross-checking our
calculations, and Jim Annis for providing the galaxy redshift 
distribution from DES mock catalogs. In addition, we are grateful to 
Wayne Hu and Andrew Zentner for helpful discussions.

This work was supported by the Kavli Institute for Cosmological 
Physics at the University of Chicago through grants NSF PHY-0114422 and 
NSF PHY-0551142, and by the US Department of Energy, including grants 
DE-FG02-95ER40896 and DE-FG02-92-ER40699. LH acknowledges support by the
Initiatives in Science and Engineering Program
at Columbia University.  ER was funded by the Center for Cosmology and 
Astro-Particle Physics (CCAPP) at The Ohio State
University, and by NSF grant AST 0707985.
ES is supported by DOE grant DE-AC02-98CH10886.

%%%%%%%%%%%%%%%%%%%%%%%%%%%%%%%%%%%%%%%%%%%%%%%%%%%%%%%%%%%%%%%%%%%%%%%%%%%%%
%%%%%%%%%%%%%%%%%%%%%%%%%%%%%%%%%%%%%%%%%%%%%%%%%%%%%%%%%%%%%%%%%%%%%%%%%%%%%

\appendix

%%%%%%%%%%%%%%%%%%%%%%%%%%%%%%%%%%%%%%%%%%%%%%%%%%%%%%%%%%%%%%%%%%%%%%%%%%%%%
\section{Rigorous derivation of shear corrections}
\label{app:exact}

This section presents derivations of the exact expressions for the lensing bias and
source clustering contributions to shear correlations.
To keep the expressions as general as possible, it is useful to define a 
window function $\Wth(\vx,\vx')$, where $\vx$, $\vx'$ stand for positions on 
the sky, which is normalized so that:
\be
\int d^2 x'\: \Wth(\vx,\vx') = 1.
\ee
In the case of correlation functions, $\Wth$ picks out galaxies separated 
by $\theta\pm d\th$. More generally, other shear observables such as top-hat 
variance or aperture mass can also be written in this way. Again, we
write the un-pixelized estimators for the shear-shear correlation, 
$\hat\xi_{ab}(\th)$ [\refeq{xihat1}],
and the galaxy-shear correlation, $\hat\xi_{ga}(\th)$ (see \refsec{galshear}) as:
\bea
\hat\xi_{ab}(\th) &=&  \frac{1}{N} \sum_{ij}\Wth(i,j)\,\nobs(i) d\Omega\, \gamma_a(i)\,\nobs(j) d\Omega\, \gamma_b(j)\label{eq:xihatA} \\
 & & N = \sum_{ij} \Wth(i,j)\,\nobs(i)d\Omega\,\nobs(j)d\Omega,\\
\hat\xi_{ga}(\th) &=&  \sum_{ij}\Wth(i,j)\,\frac{\nobs(i)}{N'(i)} d\Omega\, \gamma_a(i)\,\frac{\nfgobs(j)-\nbar^{\rm fg}}{\nbar^{\rm fg}} d\Omega, \\
 & & N'(i) = \sum_{j} \Wth(i,j)\,\nobs(j)d\Omega. \label{eq:xihatB}
\eea
We now take the infinitesimal patches of \refeqs{xihatA}{xihatB} to the 
continuum limit, so that $\nobs(i)$ becomes $\nbar [1 + \dobs(\vx)]$, 
and write the estimators for galaxy-shear and shear-shear correlations as integrals: 
\bea
\hat \xi_{ab}(\th) &=& \frac{\int d^2x \int d^2x' 
[1+\dobs(\vx)] \g_a(\vx)\: [1+\dobs(\vx')] \g_b(\vx')\:  \Wth(\vx,\vx')}
{\int d^2x'' \int d^2x''' [1+\dobs(\vx'')] [1+\dobs(\vx''')] 
\Wth(\vx'',\vx''')} \\
\hat \xi_{ga}(\th) &=& \int \frac{d^2x}{\Delta\Omega} \int d^2x' \dobs^{\rm fg}(\vx)
\frac{[1+\dobs(\vx')] \g_a(\vx')}{\int d^2x'' [1+\dobs(\vx'')] \Wth(\vx,\vx'')} \Wth(\vx,\vx')
\eea
The exact expressions for the expectation values of these estimators are given 
by:
\bea
\<\hat \xi_{ab}(\th)\> &=& \int \frac{d^2x}{\Delta\Omega} \int d^2x' 
\<  \frac{ [1+\dobs(\vx)] \g_a(\vx)\: [1+\dobs(\vx')] \g_b(\vx')}
{\int d^2x''/\Delta\Omega\int d^2x''' [1+\dobs(\vx'')] [1+\dobs(\vx''')] \Wth(\vx'',\vx''')} 
\> \:  \Wth(\vx,\vx')\\
\<\hat \xi_{ga}(\th)\> &=& \int \frac{d^2x}{\Delta\Omega} \int d^2x' 
\< \frac{\dobs^{\rm fg}(\vx)\:
[1+\dobs(\vx')] \g_a(\vx')}{\int d^2x'' [1+\dobs(\vx'')] \Wth(\vx,\vx'')} 
\> \Wth(\vx,\vx')
\eea
Now we neglect the integrals outside of the correlators, which 
essentially smooth the correlation
functions over the separations defined by the angular bin width.
Using some additional notation, we can then write the expectation values of the
quadratic estimators as follows:
\bea
\<\hat \xi_{ab}(\th)\> &=& \<  \frac{ [1+\dobs(1)] \g_a(1)\:
[1+\dobs(2)] \g_b(2)}
{1 + 2\overline{\dobs} + \widehat{\dobs\dobs}} \>\label{eq:xiQkkexp}\\
\<\hat \xi_{ga}(\th)\> &=& \< \frac{\dobs^{\rm fg}(1)\:
[1+\dobs(2)] \g_a(2)}{1+\widetilde{\dobs}(1)} \>, \label{eq:xiQgkexp}
\eea
where we have defined:
\bea
\widetilde{\dobs}(\vx) &\equiv& \int d^2x'\:\dobs(\vx')\:\Wth(\vx,\vx')\\
\overline{\dobs} &\equiv& \int \frac{d^2x}{\Delta\Omega} \:\dobs(\vx)\\
\widehat{\dobs\dobs} &\equiv& \int \frac{d^2x}{\Delta\Omega} \int d^2x'
\dobs(\vx)\dobs(\vx')\:\Wth(\vx,\vx')\label{eq:wdd}
\eea
The first quantity is the observed overdensity averaged over an annulus
around the given location. Hence, it is evaluated at position `1', giving
the overdensity of observed galaxies in an annulus around that position. 
For sufficiently large separations $\th$, $\widetilde{\dobs}$ will be small,
while it will be of order unity for separations close to the galaxy correlation
length. $\overline{\dobs}$ is the overdensity of the galaxy sample (including
magnification) averaged over the 
whole survey, measured relative to the ensemble average or an
infinite-volume survey. We have not neglected this averaged overdensity
for the sake of completeness, although for actual wide-field surveys this will be
negligible. Finally, $\wdd$ is a product of 
overdensities smoothed over separations around $\th$. Note that this
quantity is within the expectation value and hence cannot immediately be replaced
by $\xi_{gg}(\th)$.

We now expand the expectation values of the correlation functions up to 
fourth order in $\d$ and $\k$. For sake of brevity, we keep $\dobs$ [\refeq{dobs}]
unexpanded:
\bea
\<\hat \xi_{ab}(\th)\> &=& \< \g_a(1)\g_b(2) \> \nonumber \\
& & + \< \left [ \dobs(1) - \udobs \right ] \g_a(1)\g_b(2) \> + \< \g_a(1) \left [ \dobs(2) - \udobs \right ] \g_b(2) \> \nonumber \\
& & + \< \left [\dobs(1)\dobs(2) - \wdd\right ] \g_a(1)\g_b(2)\> \nonumber\\
& & - 2 \<\udobs \left [\dobs(1)+\dobs(2)\right ] \g_a(1)\g_b(2) \>
+ 2 \< \udobs^2 \g_a(1) \g_b(2) \> \label{eq:biasQkk}\\
\<\hat \xi_{ga}(\vartheta)\> &=& \< \dobs^{\rm fg}(1)\g_a(2) \> \nonumber\\
& & + \< \dobs^{\rm fg}(1) \left [\dobs(2) - \tdobs(1)\right ] \g_a(2) \> \nonumber\\
& & + \< \dobs^{\rm fg}(1) \left [\tdobs^2(1) - \tdobs(1) \dobs(2) \right ] \g_a(2) \> \label{eq:biasQgk}
\eea
To be explicit, the term in the third line for $\xi_{ab}$ stands for:
\be
\int d^2x_1\int d^2x_2\:\Wth(\vx_1,\vx_2) \<
\left [ \dobs(\vx_1)\:\dobs(\vx_2) - \int d^2x_3 \int d^2x_4\:\Wth(\vx_3,\vx_4)
\:\dobs(\vx_3)\:\dobs(\vx_4) \right ] \g_a(\vx_1)\:\g_b(\vx_2) \>
\ee

Note that the form of all magnification/size bias corrections in the final
expressions correspond to a replacement of the form:
\be
\g(1) \rightarrow \g(1) \left \{ 1 + \dobs(1) - S[\dobs](1) \right \},\ \mbox{etc.,}
\ee
where $S[\dobs]$ is some smoothing of $\dobs$ over annuli or the
whole survey. 
Hence, all corrections vanish if either, $\dobs = S[\dobs]$, i.e.
there is no observed clustering of background galaxies (including magnification
effects), or if the shear field $\g$ is uncorrelated with
$\dobs$. This holds analogously for a pixelized estimator (see \refapp{pixelest})
and is in line with 
the intuititive understanding of the magnification corrections.
Keeping only the cubic terms of \refeq{biasQkk}, and neglecting $\overline{\dobs}$,
we arrive at \refeq{xiexp}. See Appendix~\ref{app:projection} for a
discussion of the four-point terms.

%%%%%%%%%%%%%%%%%%%%%%%%%%%%%%%%%%%%%%%%%%%%%%%%%%%%%%%%%%%%%%%%%%%%%%%%%%%%%
\section{Pixel-based shear estimators}
\label{app:pixelest}

An alternative approach to estimating shear correlations is to divide
the survey volume into pixels $\a$ of finite volume defined by window functions 
$\W_\a(\vx)$ (normalized so that $\int \W_\a(\vx)\,d^2 x=1$). Each pixel contains many galaxies, and
one estimates the shear and galaxy overdensity 
directly for each pixel (e.g., \cite{HuWhite,MunshiEtal}; again setting 
all weights to 1): 
\bea
\ghat_a(\a) &=& 
\frac{1}{n_{\rm pix}(\a)}\sum_i n_{\rm obs}(i) \:\gamma_a(i)\: \W_\a(i)\label{eq:ghat}\\
\dhat(\a) &=& \frac{n^{\rm fg}_{\rm pix}(\a)}{\bar{n}^{\rm fg}_{\rm pix}} - 1 = 
\sum_i \frac{n^{\rm fg}_{\rm obs}(i) - \bar n^{\rm fg}}{\bar n^{\rm fg}} \W_\a(i) \label{eq:dhat}
\eea
Here, $\gamma_a(i)$ is the shear measured from galaxy $i$, $\nobs^{\rm fg}$ is
the {\it foreground} galaxy density, and we have again subdivided
the finite-sized pixel into infinitesimal patches $i$, so that the 
observed number of galaxies $\nobs(i)$ in each patch is either 0 or 1. 
$n_{\rm pix}(\a)$ is the number of galaxies 
observed in pixel $\a$, while $\bar n_{\rm pix}$ is the expected average 
number of galaxies per pixel:
\be
n_{\rm pix}(\a) = \sum_i n_{\rm obs}(i) \W_\a(i),\quad
\bar n_{\rm pix} = \sum_i \nbar \W_\a(i) = \nbar,
\ee
and analogously for the foreground galaxy densities.
The estimators \refeq{ghat} and (\ref{eq:dhat}) result in pixelized maps
of the shear components and foreground galaxy overdensities, which can then be
processed in real or Fourier space to measure shear correlations.

Going to the continuum limit of \refeq{ghat} and (\ref{eq:dhat}) yields:
\bea
\ghat_a(\a) &=& \frac{\int d^2 x \:[1 + \dobs(\vx)] \g_a(\vx)\: \W_\a(\vx)}
{\int d^2 x'\: [1 + \dobs(\vx')]\: \W_\a(\vx')}\\
\dhat(\a) &=& \int d^2x \frac{\nobs^{\rm fg}(\vx)-\nbar^{\rm fg}}{\nbar^{\rm fg}} \W_\a(\vx), \quad \nbar^{\rm fg} = \int \frac{d^2 x}{\Delta\Omega} \nobs(\vx)
\eea
We now take expectation values of correlators of the pixelized shear
and overdensity fields, neglecting the effect of pixelization on the
correlation function (which is appropriate if the separation $\th$ is much larger
than the pixel scale):
\bea
\<\hat \xi_{ab}(\theta)\> &=& \< 
\frac{\overline{(1+\dobs) \g_a}(1)}{1 + \overline{\dobs}(1)}
\frac{\overline{(1+\dobs) \g_b}(2)}{1 + \overline{\dobs}(2)} \>
\label{eq:xiLkkexp}\\
\<\hat \xi_{ga}(\theta)\> &=& \< \overline{\dobs^{\rm fg}}(1)\: 
\frac{\overline{(1+\dobs) \g_a}(2)}{1 + \overline{\dobs}(2)} \>,
\label{eq:xiLgkexp}
\eea
where we have labeled two points separated by $\theta$ as 1 and 2.
Here, barred quantities denote averages over pixels:
\be
\overline{X}(\a) \equiv \int d^2 x'\: X(\vx') \W_\a(\vx').
\ee
In \refeq{xiLgkexp}, we have neglected a correction due to the integral
constraint \cite{HuiGazta}, since $\nbar$ is measured in the survey itself.
However, this effect is of order the overdensity averaged over the whole
survey, and hence very small for large surveys. In contrast, the denominators
kept in \refeqs{xiLkkexp}{xiLgkexp} are integral constraints which are 
important, since
they are of order of the overdensity averaged over \textit{pixel} scales.

Expanding the expectation values to fourth order, we obtain for the
pixelized estimators:
\bea
\<\hat \xi_{ab}(\theta)\> &=& \<\bg_a(1)\bg_b(2)\> \nonumber \\
& & + \<\bg_a(1) \left [ \overline{\dobs\g_b}(2) - \bdobs(2)\bg_b(2) \right ] \> 
+ \<\left [ \overline{\dobs\g_a}(1) - \bdobs(1)\bg_a(1) \right ] \bg_b(2) \> \nonumber\\
& & + \< \left [ \overline{\dobs\g_a}(1) - \bdobs(1)\bg_a(1) \right ]\:
 \left [ \overline{\dobs\g_b}(2) - \bdobs(2)\bg_b(2) \right ] \> \nonumber \\
& & + \< \left [\bdobs^2(1)\bg_a(1) - \bdobs(1)\overline{\dobs\g_a}(1) \right ] \bg_b(2) \>
+ \< \bg_a(1) \left [\bdobs^2(2)\bg_b(2) - \bdobs(2)\overline{\dobs\g_b}(2) \right ] \> \label{eq:xiLexp}\\
\<\hat \xi_{ga}(\theta)\> &=& \<\bdobsfg(1)\bg_a(2)\>\nonumber \\
& & + \< \bdobsfg(1) \left [ \overline{\dobs\g_a}(2) - \bdobs(2)\bg_a(2) \right ] \> \nonumber\\
& & + \< \bdobsfg(1) \left [ \bdobs^2(2)\bg_a(2) - \bdobs(2)\overline{\dobs\g_b}(2) \right ] \> \\
\eea
Clearly, all corrections vanish if $\overline{\dobs\,\g} = \bdobs\,\bg$
within a correlator, which is the case if either the observed galaxy 
distribution $\dobs$ {\it or} the shear $\g$ are smooth on pixel scales,
or if they are completely uncorrelated.

For the same reasons as for the unpixelized estimator, detailed in
\refapp{projection}, the quartic corrections are much smaller than
the cubic corrections. Repeating the derivation leading to \refeq{dCkappa},
and noting that in Fourier space the smoothed convergence field is given
by:
\be
\bk(\vl) = \Wt(\vl)\:\k(\vl),\quad \Wt(\vl) \equiv \int d^2 x\:W_\alpha(\v{x})
e^{-i \vl\cdot\v{x}},
\ee
we obtain the following expression for the magnification correction to the 
shear power $C^\k(\l)$ in case of the pixelized estimator:
\bea
\Delta C_\l &=& 2q
\int \frac{d^2\l_1}{(2\pi)^2} \cos 2\phi_{\l_1}
\left[ \Wt(\vl) - \Wt(\vl_1) \Wt(-\vl-\vl_1) \right] 
B(\vl,\vl_1,-\vl - \vl_1)\\
&\approx& 2q
\int \frac{d^2\l_1}{(2\pi)^2} \cos 2\phi_{\l_1}
\left[ 1 - |\Wt(\vl_1)|^2 \right] 
B(\vl,\vl_1,-\vl - \vl_1) \label{eq:dCkappaL}.
\eea
For the second approximate equality, we have assumed that $\l \ll 1/\th_{\rm pix}$,
where $\th_{\rm pix}$ is the angular size of the pixels, so that 
$\Wt(\vl)\approx 1$. The factor in square
brackets in \refeq{dCkappaL} is the only difference in the magnification 
correction for the pixelized estimator compared to \refeq{dCkappa}.
This factor acts as a high pass, so that
only modes with $\l \gtrsim 1/\th_{\rm pix}$ contribute to the magnification
corrections, as expected from \refeq{xiLexp} and our discussion above. Hence,
for sufficiently small pixels, the magnification corrections are suppressed.

Note, however, that our derivations assumes that in estimating shear correlations from pixelized
maps, all pixels
receive the same weight. If a weighting scheme according to signal-to-noise
is used, weighting each pixel by the number of observed galaxies within the pixel (as
appropriate for inverse variance weighting), the
magnification correction is re-introduced, and we essentially go back to
\refeq{dCkappa}. Note that in any case the reduced shear correction is not 
suppressed by using small pixels, and is always given by \refeq{dCkappa}.

%%%%%%%%%%%%%%%%%%%%%%%%%%%%%%%%%%%%%%%%%%%%%%%%%%%%%%%%%%%%%%%%%%%%%%%%%%%%%
\section{Quartic corrections} 
\label{app:projection}

Going from \refeq{biasQkk} to \refeq{xiexp}, we have neglected
several 4-point correlations. For the estimators we consider, all of
the 4-point terms due to lensing bias are suppressed by roughly two 
orders of magnitude compared to
the cubic terms, as we dicuss below. Similar conclusions hold for 
the 4-point terms in the pixelized estimator. In the following, we
again neglect the galaxy overdensity averaged over the whole survey,
$\overline{\dobs}$.

The 4-point contributions can be divided into 
three classes. First, there are source-lens clustering terms:
\bea
\Delta\<\hat \xi_{ab}(\th)\>_{\rm quartic,I} & = & \< [\d_g(1)\d_g(2) - \wddg ] \g_a(1)\g_b(2) \>_{\rm connected} \\
& + & \< \d_g(1)\g_a(1)\> \<\d_g(2)\g_b(2)\> - \<\wdh{\d_g\g_a(1)\rangle\langle\d_g}\g_b(2)\> \\
& + & \< \d_g(1)\g_b(2)\> \<\d_g(2)\g_a(1)\> - \<\wdh{\d_g\g_a(1)\rangle\langle\d_g}\g_b(2)\> \\
& + & \< \d_g(1)\d_g(2)\> \<\g_a(1)\g_b(2)\> - \<\wddg\>\<\g_a(1)\g_b(2)\>.
\eea
The first, connected term is related to the matter four-point function.
The second and third lines contain the actual source-lens terms.
Note that for the second terms in each line, 
the product of the two correlators is to be integrated over
following \refeq{wdd}.
The two terms in the fourth line cancel, since $\<\wddg\> = \<\d_g(1)\d_g(2)\>$
(see \refeq{wdd}). This cancelation is a consequence of the normalizing denominator
in \refeq{xihatA}: when measuring the shear, we divide by the
number of pairs of observed galaxies with given separation $\th$.

The second set of quartic contributions are mixed source-lens clustering
and lensing bias terms:
\bea
\Delta\<\hat \xi_{ab}(\th)\>_{\rm quartic,II}  & = & 
q \< [\d_g(1)\k(2) - \wddkg ] \g_a(1)\g_b(2) \>_{\rm connected} \\
& + & q \left \{ \< \d_g(1)\g_a(1)\> \<\k(2)\g_b(2)\> - \<\wdh{\d_g\g_a(1)\rangle\langle\k}\g_b(2)\> \right \} \\
& + & q \left \{ \< \d_g(1)\g_b(2)\> \<\k(2)\g_a(1)\> - \<\wdh{\d_g\g_a(1)\rangle\langle\k}\g_b(2)\> \right \} \\
& + & q \left \{ \< \d_g(1)\k(2)\> \<\g_a(1)\g_b(2)\> - \<\wddkg\>\<\g_a(1)\g_b(2)\> \right \} \\
& + & \{ \d_g(1)\k(2) \leftrightarrow \k(1)\d_g(2) \}.
\eea
As expected, these are all proportional to $q$. The first three lines again
give the contributing source-lens clustering/lensing bias contributions,
while the terms in the fourth line cancel.

Finally, the quartic terms from ``pure'' lensing bias receive two contributions:
first, from the quartic terms in \refeq{biasQkk}. Second, there are 
quadratic contributions to $\dobs$ from lensing bias. Expanding the
lensing magnification $A = [(1-\k)^2 - |\g|^2]^{-1/2}$ to second
order, we obtain \cite{SchmidtEtal}\footnote{We neglect second derivatives
of $\nobs$ with respect to $\ln f$, $\ln r$ here.}:
\be
\dobs = \d_g + q\,\k + c_1\,\k^2 + c_2\,|\g|^2,
\ee
where $c_1 =q(q+1)/2$, $c_2=q/2$, and $|\g|^2 = \g_1^2 + \g_2^2$. Together,
we obtain the following quartic terms due to lensing bias:
\bea
\Delta\<\hat \xi_{ab}(\th)\>_{\rm quartic,III}  & = & 
q^2 \< [\k(1)\k(2) - \wddk ] \g_a(1)\g_b(2) \>_{\rm connected} \label{eq:kla} \\
& + & 2\,q^2 \left \{ \< \k(1)\g_a(1)\> \<\k(2)\g_b(2)\> - \<\wdh{\k\g_a(1)\rangle\langle\k}\g_b(2)\> \right \} \label{eq:klb} \\
& + & q \left \{ \< \k(1)\k(2)\> \<\g_a(1)\g_b(2)\> - \<\wddk\>\<\g_a(1)\g_b(2)\> \right \} \label{eq:klc} \\
& + &  c_1 \<\k^2(1)\g_a(1)\g_b(2)\>_{\rm connected} +
c_2 \<|\g(1)|^2\g_a(1)\g_b(2)\>_{\rm connected} + \{ (1) \leftrightarrow (2) \} \label{eq:kld} \\
& + & \left \{ c_1 \<\k^2(1)\> + c_2 \< |\g(1)|^2\> \right \} \<\g_a(1)\g_b(2)\> 
+ \{ (1) \leftrightarrow (2) \} \label{eq:kle} \\
& + & 2\,c_1 \<\k(1)\g_a(1)\> \<\k(1)\g_b(2)\> + 
2\,c_2 \sum_{c=1,2} \<\g_c(1)\g_a(1)\> \<\g_c(1)\g_b(2)\> 
+ \{ (1) \leftrightarrow (2) \}.\label{eq:klf}
\eea
Here, $\{ (1) \leftrightarrow (2) \}$ means that $\k^2(1)$, $|\g(1)|^2$ are
to be replaced with $\k^2(2)$, $|\g(2)|^2$, respectively.
Line~\ec{kla} and \ec{kld} are connected terms given by the shear four-point
function. We will discuss those below. The terms in line~\ec{klc} cancel
in the same way as the corresponding source-lens terms. The terms in lines~\ec{klb},
\ec{kle}, \ec{klf} are proportional to $\xi_{\k\k}(\th)^2$, or
$\xi_{\k\k}(\th)\,\xi_{\k\k}(0)$. In other words, the relative magnitude of these
corrections is of order $\xi_{\k\k}(0)\sim $few~$10^{-4}$ or less. Hence,
we can safely neglect them compared with the percent-level of the cubic
corrections.

\begin{figure}[t]
 %\begin{center}
\begin{minipage}[t]{0.48\textwidth}
\includegraphics[width=\textwidth]{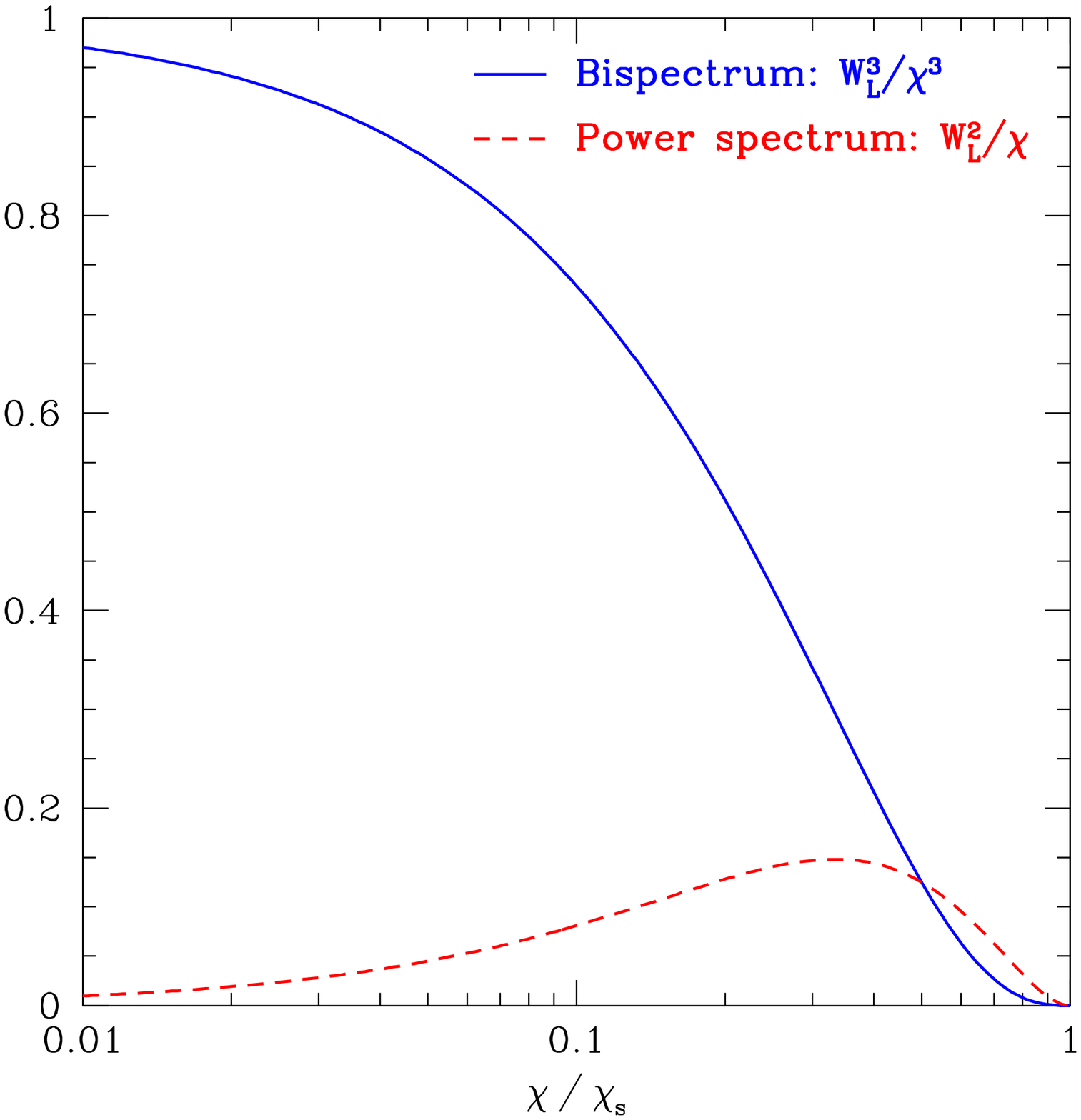}
\end{minipage}
\hfill
\begin{minipage}[t]{0.48\textwidth}
\includegraphics[width=\textwidth]{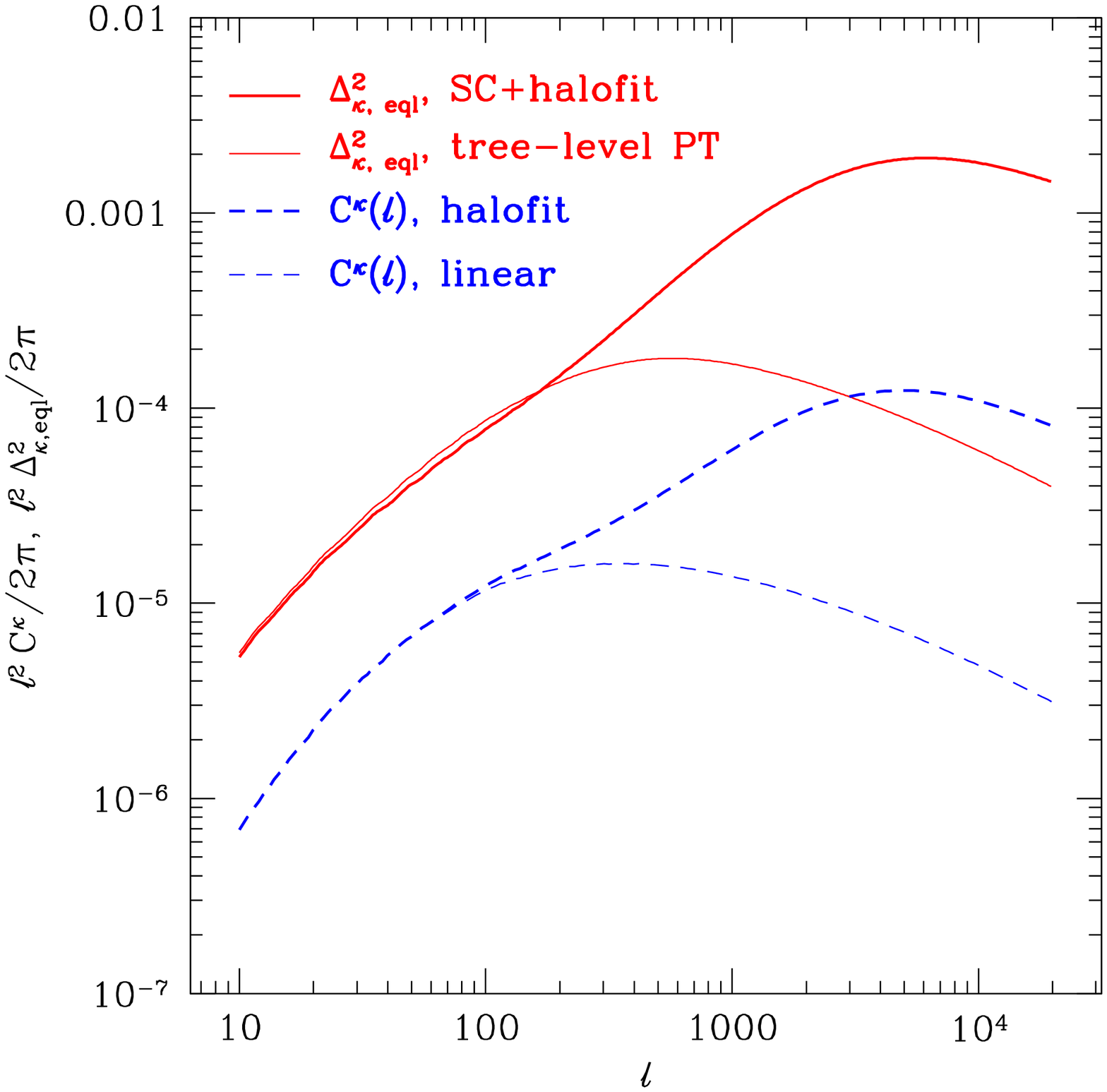}
\end{minipage}
%\end{center}
\caption{\textit{Left panel:} Lensing weight functions for the shear power
spectrum, $W_L(\chi_s,\chi)^2/\chi$ (red, dashed), and the shear bispectrum, 
$W_L(\chi_s,\chi)^3/\chi^3$ (blue, solid), in units of $\chi_s$.
\textit{Right panel:} Scaled shear power $\l^2/(2\pi)\:C^\k(\l)$ (blue, dashed),
and equilateral bispectrum power $\l^2/(2\pi)\:\sqrt{B^\k(\l,\l,\l)}$ (red,
solid), for $z_s=1$. The thin lines show the linear/tree-level prediction, 
while the thick lines are using the non-linear fitting formulas of
\cite{SmithEtal} and \cite{SC}. 
\label{fig:D2kappa}}
\end{figure}

In order to understand why the cubic corrections are so much more important,
consider the expressions for the shear power spectrum (2-point function) and 
bispectrum (3-point function).
Using the Limber or small-angle approximation, the shear power for sources
at a fixed redshift $z_s$, with $\chi_s=\chi(z_s)$, can be written
as a projection of the matter power spectrum $P(k,\chi)$:
\be
C^\k(\l) = \left(\frac{3}{2}\Omega_m\:H_0^2 \right)^2 \int_0^{\chi_s}
\frac{d\chi}{\chi} \frac{W_L(\chi_s,\chi)^2}{\chi \: a^2(\chi)}  P(\l/\chi;\chi).
\label{eq:Ckappa}
\ee
Here, $\chi$ denotes comoving distance, 
$W_L(\chi_s,\chi)=\chi/\chi_s(\chi_s-\chi)$, and $a$ is the scale factor.
Similarly, the shear bispectrum $B^\k$ can be written as a projection of
the matter bispectrum $B(k_1,k_2,k_3;\chi)$:
\be
B^\k(\vl_1,\vl_2,\vl_3) = \left(\frac{3}{2}\Omega_m\:H_0^2 \right)^3 \int_0^{\chi_s}
\frac{d\chi}{\chi} \left ( \frac{W_L(\chi_s,\chi)}{\chi\:a(\chi)} \right )^3
B \left (\frac{\vl_1}{\chi},\frac{\vl_2}{\chi},\frac{\vl_3}{\chi}; \chi \right ),
\label{eq:Bkappa}
\ee
In case the sources are distributed according to a broad redshift distribution, 
$dN/dz$ (assumed normalized
to unity), $W_L$ in \refeqs{Ckappa}{Bkappa} is to be replaced with:
\be
W_{L,dN/dz}(\chi) = \frac{1}{H(\chi)}\int_{z(\chi)}^{\infty} dz_s W_L(\chi(z_s),\chi)\: \frac{dN}{dz}(z_s).
\ee
Now, for the 3D matter field, $\<\d(1)\d(2)\d(3)\>$ is of the same
order of magnitude as
$\<\d(1)\d(2)\>\<\d(2)\d(3)\> + $cycl.. However, in case of the shear, which is
proportional to the {\it projected} density field, $B^\k(\l,\l,\l)$ is
larger than $C^\k(\l)^2$ by a factor of order several hundreds (right panel
of \reffig{D2kappa}).
This is because the
two-point and three-point functions are projected with a different weighting 
of low-$z$ contributions. \refFig{D2kappa} (left panel)
shows the effective weight functions for $C^\k$ (red, dashed) and $B^\k$ 
(blue, solid). Clearly,
the late-time contributions receive more weight in case of the bispectrum, which grows
as $\sim D(a)^4$. In addition, the low-$z$ contributions along the line
of sight are probed at smaller scales, which additionally enhances the bispectrum.
For this reason, the cubic corrections dominate the disconnected quartic 
contributions in \refeqs{kla}{klf},
even though they are formally of the same order in perturbation theory.

The connected four-point terms, \refeq{kla} and \ec{kld}, are given
by the convergence trispectrum.
We can roughly estimate the contribution from these terms relative
to the cubic terms as:
\be
\frac{\Delta C^\k_{\rm quartic}(\l)}{\Delta C^\k_{\rm cubic}(\l)} \sim
\frac{\l^4\:T^\k_{\rm sq}(\l)}{\l^2\: B^\k_{\rm eql}(\l)} = 
2\pi \frac{(\Delta^2_{\rm sq})^3}{(\Delta^2_{\rm eql})^2} \lesssim 0.05
\;\;\mbox{for}\;\l \leq 10^4.
\label{eq:4pt}
\ee
Here, $T^\k_{\rm sq}$ denotes the square trispectrum, $B^\k_{\rm eql}$ denotes
the equilateral bispectrum, and the scaled quantities $\Delta^2_{\rm eql}$,
$\Delta^2_{\rm sq}$ were defined and calculated in
\cite{CoorayHuBi,CoorayHuTri}:
\be
\Delta^2_{\rm eql}(\l) \equiv \frac{\l^2}{2\pi} [B^\k_{\rm eql}(\l)]^{1/2},\quad
\Delta^2_{\rm sq}(\l) \equiv \frac{\l^2}{2\pi} [T^\k_{\rm sq}(\l)]^{1/3}.\quad
\ee
Note that for very small scales, or in case \refeq{4pt} underestimates the 
size of the connected (non-Gaussian) quartic
contributions, the quartic contribution
in \refeq{kla}, \ec{kld} are positive and act to increase the 
magnification corrections.

\section{B-Modes}
\label{app:b}

The new terms induced by lensing bias produce B-modes in addition to the E-modes. This is somewhat akin to the well-known effect of lensing of the cosmic microwave background, when large scale structure distorts the E-modes. Cooray and Hu~\cite{Cooray:2002mj} examined the B-modes induced by higher order corrections to the Born approximation. Lensing bias (and reduced shear) lead to an additional source of B-modes. The estimator for the B-mode is:
\begin{equation}
\hat B(\vl) = \sin(2\phi_\l) \gamma^{\rm obs}_1(\vl) - \cos(2\phi_\l)\gamma^{\rm obs}_2(\vl)
\end{equation}
with associated power spectrum:
\begin{eqnarray}
C^B(\l) &=& \int \frac{d^2\l'}{(2\pi)^2} \langle \hat B(\vl) \hat B(\vl')\rangle\vs 
&=& \int \frac{d^2\l'}{(2\pi)^2} \langle \left[\sin(2\phi_\l) \gamma^{\rm obs}_1(\vl) - \cos(2\phi_\l)\gamma^{\rm obs}_2(\vl)\right]
\left[\sin(2\phi_{\l'}) \gamma^{\rm obs}_1(\vl') - \cos(2\phi_{\l'})\gamma^{\rm obs}_2(\vl')\right]\rangle.\eql{bmode}
\end{eqnarray}
In the absence of lensing corrections, $\gamma^{\rm obs}_1(\vl)=\cos(2\phi_\l)\kappa(\vl)$ and $\gamma^{\rm obs}_2(\vl)=\sin(2\phi_\l)\kappa(\vl)$, so the power spectrum vanishes identically. Lensing effects lead to a new term when the shears are estimated:
\begin{equation}
\gamma^{\rm obs}_a(\v{x}) \rightarrow \gamma_a(\v{x})\left[ 1+ (1+q)\kappa(\v{x})\right]
\end{equation}
or in Fourier space,
\newcommand\vq{\vec{q}}
\begin{equation}
\gamma^{\rm obs}_a(\vl) \rightarrow \gamma_a(\vl) + (1+q) \int \frac{d^2\l'}{(2\pi)^2} \gamma_a(\vl')\kappa(\vl-\vl').
\end{equation}
The second term here is the only one that survives when computing the B-mode spectrum.
Inserting these into \refeq{bmode} leads to
\begin{eqnarray}
C^B(\l) &=& (1+q)^2 \int \frac{d^2\l'}{(2\pi)^2} \int\frac{d^2\l''}{(2\pi)^2}\int\frac{d^2\l'''}{(2\pi)^2} \langle \left[\sin(2\phi_\l) \cos(2\phi_{\l''}) \kappa(\vl'') \kappa(\vl-\vl'')  - \cos(2\phi_\l)\sin(2\phi_{\l''}) \kappa(\vl'') 
\kappa(\vl-\vl'')\right]\vs&&\times
\left[\sin(2\phi_{\l'}) \cos(2\phi_{\l'''}) \kappa(\vl''') \kappa(\vl'-\vl''') - \cos(2\phi_{\l'})\sin(2\phi_{\l'''}) \kappa(\vl''')\kappa(\vl'-\vl''')\right]\rangle\vs 
&=& (1+q)^2
\int \frac{d^2\l'}{(2\pi)^2} \int\frac{d^2\l''}{(2\pi)^2}\int\frac{d^2\l'''}{(2\pi)^2} 
\sin(2\phi_\l-2\phi_{\l''}) \sin(2\phi_{\l'}-2\phi_{\l'''})
\langle \kappa(\vl'') \kappa(\vl-\vl'')\kappa(\vl''') \kappa(\vl'-\vl''')\rangle.\eql{conv}
\end{eqnarray}
Apart from the $l=0$ mode, there are two ways to contract the (assumed) Gaussian convergence fields in \refeq{conv} 
%{\bf FS: added trispectrum from my notes}:
\be
\langle \kappa(\vl'') \kappa(\vl-\vl'')\kappa(\vl''') \kappa(\vl'-\vl''')\rangle
%&=& (2\pi)^2 \delta^2(\vl+\vl') \left\{ T^\k(\vl'',\vl-\vl'',\vl''',-\vl-\vl''') \right .\vs
= (2\pi)^4 \delta^2(\vl+\vl') C^\kappa(\l'') C^\kappa(\vert\vl-\vl''\vert) 
\left[ \delta^2(\vl''+\vl''') + \delta^2(\vl''-\vl-\vl''')\right] ,
\ee
so, choosing $\phi_\l=0$ leads to
\begin{equation}
C^B(\l) = (1+q)^2 \int\frac{d^2\l'}{(2\pi)^2}
\sin(2\phi_{\l'}) 
C^\kappa(\l') C^\kappa(\vert\vl-\vl'\vert) \left[ \sin(2\phi_{\l'})+ \sin(2\phi_{\vl'-\vl})\right].
\end{equation}
Apart from geometric factors, this is of order $\l^2C^\kappa(\l) \sim 10^{-4}$ 
smaller than the E-mode spectrum, in qualitative agreement with the terms analyzed in \cite{Cooray:2002mj}. Note that here the trispectrum terms may contribute an even larger correction on small scales. We leave this calculation for future work.

%%%%%%%%%%%%%%%%%%%%%%%%%%%%%%%%%%%%%%%%%%%%%%%%%%%%%%%%%%%%%%%%%%%%%%%%%%%%%
%%%%%%%%%%%%%%%%%%%%%%%%%%%%%%%%%%%%%%%%%%%%%%%%%%%%%%%%%%%%%%%%%%%%%%%%%%%%%
\bibliography{sizebias}

\end{document}